\documentclass[a4paper]{article}

\usepackage{icrc2013}
\usepackage[english]{babel}

\title{Energy spectra of KASCADE-Grande based on shower size measurements and different hadronic interaction models}

\shorttitle{Energy spectra of KASCADE-Grande}

\authors{
D. Kang$^{1}$, 
W.D. Apel$^{2}$, 
J.C. Arteaga-Vel\'azquez$^{3}$,
K. Bekk$^{2}$, 
M. Bertaina$^{4}$,
J. Bl\"umer$^{1,2}$, 
H. Bozdog$^{2}$,
I.M. Brancus$^{5}$, 
E. Cantoni$^{4,6,a}$, 
A. Chiavassa$^{4}$,
F. Cossavella$^{1,b}$,
C. Curcio$^{4}$,
K. Daumiller$^{2}$,
\\
V. de Souza$^{7}$,
F. Di Pierro$^{4}$,
P. Doll$^{2}$,
R. Engel$^{2}$,
J. Engler$^{2}$,
B. Fuchs$^{1}$,
D. Fuhrmann$^{8,c}$,
H.J. Gils$^{2}$,
\\
R. Glasstetter$^{8}$,
C. Grupen$^{9}$,
A. Haungs$^{2}$,
D. Heck$^{2}$,
J.R. H\"orandel$^{10}$,
D. Huber$^{1}$,
T. Huege$^{2}$,
\\
K.-H. Kampert$^{8}$,
H.O. Klages$^{2}$,
K. Link$^{1}$,
P. Luczak$^{11}$,
M. Ludwig$^{1}$,
H.J. Mathes$^{2}$,
H.J. Mayer$^{2}$,
\\
M. Melissas$^{1}$,
J. Milke$^{2}$,
B. Mitrica$^{5}$,
C. Morello$^{6}$,
J. Oehlschl\"ager$^{2}$,
S. Ostapchenko$^{2,d}$,
\\
N. Palmieri$^{1}$,
M. Petcu$^{5}$,
T. Pierog$^{2}$,
H. Rebel$^{2}$,
M. Roth$^{2}$,
H. Schieler$^{2}$,
S. Schoo$^{2}$,
F. Schr\"oder$^{2}$,
\\
O. Sima$^{12}$,
G. Toma$^{5}$,
G.C. Trinchero$^{6}$,
H. Ulrich$^{2}$,
A. Weindl$^{2}$,
J. Wochele$^{2}$,
J. Zabierowski$^{11}$
\\
KASCADE-Grande Collaboration
}

\afiliations{
$^1$ Institut f\"ur Experimentelle Kernphysik, KIT - Karlsruher Institut f\"ur Technologie, Germany\\
$^2$ Institut f\"ur Kernphysik, KIT - Karlsruher Institut f\"ur Technologie, Germany\\
$^3$ Universidad Michoacana, Instituto de Fisica y Matem\'aticas, Mexico\\
$^4$ Dipartimento di Fisica, Universit\`a degli Studi di Torino, Italy\\
$^5$ National Institute of Physics and Nuclear Engineering, Bucharest, Romania\\
$^6$ Istituto di Fisica dello Spazio Interplanetario, INAF Torino, Italy\\
$^7$ Universidade S\~ao Paulo, Instituto de F\'{\i}sica de S\~ao Carlos, Brasil\\
$^8$ Fachbereich Physik, Universit\"at Wuppertal, Germany\\
$^9$ Fachbereich Physik, Universit\"at Siegen, Germany\\
$^{10}$ Dept. of Astrophysics, Radboud University Nijmegen, The Netherlands\\
$^{11}$ Soltan Institute for Nuclear Studies, {\L}\'{o}d\'{z}, Poland\\
$^{12}$ Department of Physics, University of Bucharest, Bucharest, Romania\\
\scriptsize{
$^{a}$ now at: Istituto Nazionale di Ricerca Metrologia, INRIM, Torino;\\
$^{b}$ now at: Max-Planck-Institut f\"ur Physik, M\"unchen, Germany;\\
$^{c}$ now at: University of Duisburg-Essen, Duisburg, Germany;\\
$^{d}$ now at: University of Trondheim, Norway.
}
}

\email{donghwa.kang@kit.edu}

\abstract{
KASCADE-Grande is dedicated for investigations of cosmic-ray
air showers in the primary energy range from 10 PeV to 1 EeV.
The multi-detector system allows us to reconstruct charged particles, 
electron and muon numbers for individual air showers with high accuracies.
Based on the shower size ($N_{ch}$) spectra of the charged particle component,
the all-particle energy spectrum of cosmic rays is reconstructed,
where attenuation effects in the atmosphere are corrected by applying the constant
intensity cut method.
The energy calibration is performed by using CORSIKA simulations with high-energy
interaction models QGSJET-II-2, QGSJET-II-4, EPOS 1.99 and SIBYLL 2.1, where
FLUKA has been used as low-energy interaction model for all cases.
In the different hadronic models, different abundances for shower particles are predicted.
Such model differences in the observables will be compared and discussed in this contribution. 
Furthermore, by using data with increasing statistics, 
the updated energy spectra by means of different
interaction models will be presented.}


\begin{document}
\maketitle

\section{Introduction}
Investigations of the energy spectra of elemental groups and 
mass composition of primary cosmic rays in the knee region around 10$^{17}$\ eV 
give an important clue 
to examine theoretical models of the cosmic ray origin, acceleration and propagation.
The multi-detector array of KASCADE-Grande is designed for observations
of cosmic ray air showers, in particular, in the energy range of the transition region.
Recent results of the KASCADE-Grande measurements  
have shown two spectral features in the all-particle energy spectrum \cite{bib:KG}:
a knee-like structure at 90 PeV \cite{bib:PRL} and a hardening of the spectrum \cite{bib:PRD}
around 20 PeV.
In general, interpretations of the measurements are related to
numerical simulations of extensive air showers
to obtain shower properties including the nature of the primary particles.
The relation between the observables and the primary energy depends on the
hadronic interaction models,
so that, on energy estimations, a large uncertainty in these simulations
comes from the models which describe the hadronic interactions.
In this contribution,
results from different hadronic interaction models are therefore discussed 
how their features affect the energy assignment, based on the measurements
of shower size ($N_{ch}$), i.e. the total number of charged particles.

The KASCADE-Grande experiment covering an area of about 0.5 km$^{2}$ is
optimized to measure extensive air showers up to primary energies of 1\ EeV
\cite{bib:NIM}. It consists of 37 scintillation detector stations located on a
hexagonal grid with an average spacing of 137\ m for the measurements of
electromagnetic and muonic shower components.
Each of the detector stations is equipped with
plastic scintillator sheets covering a total area of 10\ m$^{2}$.
Full efficiency
for the total number of charged particles
is reached at around 10$^{6}$, which corresponds to
a primary energy of about10$^{16}$\ eV.
The limit at high energy is due to the restricted area of the
Grande array.

\section{Hadronic interaction models}
The CORSIKA \cite{bib:Heck} program has been used for the air shower simulations,
applying different hadronic interaction models. 
High-energy interactions were used with different models of QGSJET-II-2
\cite{bib:Ostapchenko}, EPOS 1.99 \cite{bib:Pierog}, SIBYLL 2.1 \cite{bib:Ahn}.
and QGSJET-II-4\cite{bib:Ostapchenko2011}.
For hadronic interactions at low energies, the FLUKA \cite{bib:Fasso} (E $<$ 200\ GeV) model 
has been used.
The response of all detector components is taken into account by using the GEANT package.
The predicted observables at ground level, such as e.g. the number of
electrons, muons and hadrons are then compared to the measurements.

Showers induced by five different primaries (p, He, O, Si, and Fe) have been simulated. 
The simulations covered the energy range of 10$^{14}$ to 3$\times$10$^{18}$ eV
with zenith angles in the interval 0$^{\circ}$ - 42$^{\circ}$. 
The spectral index in the simulations was -2 and 
for the analysis it is weighted to a slope of -3. 
The simulated events are analyzed by the same method as the experimental data, 
in order to avoid biases by pattern recognition and reconstruction algorithms.
The systematic uncertainty of the total number of charged particles is smaller than 5\% 
and its statistical accuracy is better than 15\%.

\begin{figure}[t]
\includegraphics[width=83mm]{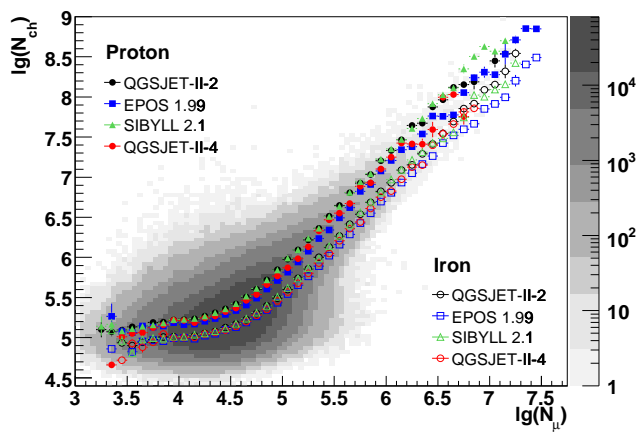}
\caption{
The 2-dimensional shower size spectrum measured by KASCADE-Grande (color-coded
area), along with proton and iron induced showers for 
QGSJET-II-2, QGSJET-II-4, EPOS 1.99 and SIBYLL 2.1 simulations.}
\label{fig1}
\end{figure}

\begin{figure}[t]
\includegraphics[width=90mm]{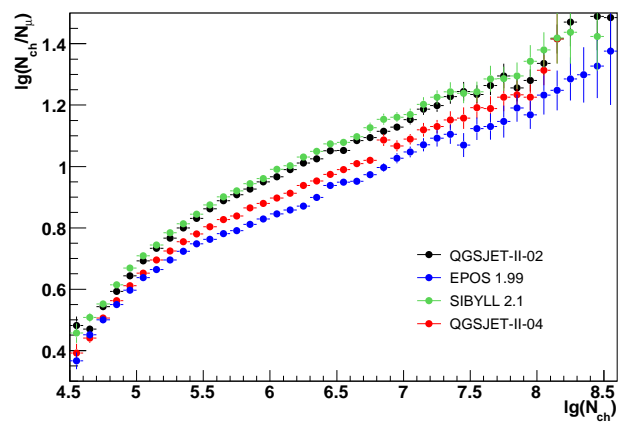}
\caption{
Relation between the number of charged particles $N_{ch}$ and 
the muon numbers $N_{\mu}$ as a function of $N_{ch}$
for four different simulations. The errors of mean values are plotted here.}
\label{fig2}
\end{figure}

\section{All-particle energy spectra}
The analysis presented here is based on the data of 1753 days with increasing statistics,
where all detector components were operating without failures in data acquisitions.
The quality cuts on the fiducial area and zenith angles smaller than 40$^{\circ}$
result in approximately 2$\cdot$10$^{7}$ events for the further analysis. 

The studies of QGSJET-II-2, EPOS 1.99 and SIBYLL 2.1 models with KASCADE-Grande data
can be found in Ref. \cite{bib:Kang}. 
Therefore, predictions of the most recent version of QGSJET-II-4 model
will be mainly investigated in this paper.

Figure \ref{fig1} represents the measured 2-dimensional shower size spectrum,
including the full detector response by simulations.
The symbols correspond to the primary protons and iron nuclei, 
as predicted by the different interaction models.
In the QGSJET-II-4 model, one obtained about 20\% enhancement 
for the shower muon content of extensive air showers,
due to the effects of the modified treatment of charge exchange processes in pion collisions.
Therefore, the most probable values for QGSJET-II-4 in Fig.\ \ref{fig1}
show a similar tendency to the EPOS 1.99 model, which has also about 15\% more muons than QGSJET-II-2 
at KASCADE-Grande energies.
This implies that a dominant light mass composition is predicted 
if the QGSJET-II-4 model is used to reconstruct the mass of primary particles
from the measured data. 
Figure \ref{fig2} shows the ratio of the number of charged particles ($N_{ch}$) 
to the muon number ($N_{\mu}$) as a function of $N_{ch}$.
Both QGSJET-II-2 and SIBYLL 2.1 models have a similar abundance ratio of $N_{ch}$ to $N_{\mu}$,
while the QGSJET-II-4 and EPOS 1.99 have approximately 10\% and 20\% more muons, respectively, 
comparing to QGSJET-II-2.

\begin{figure}[t]
\includegraphics[width=90mm]{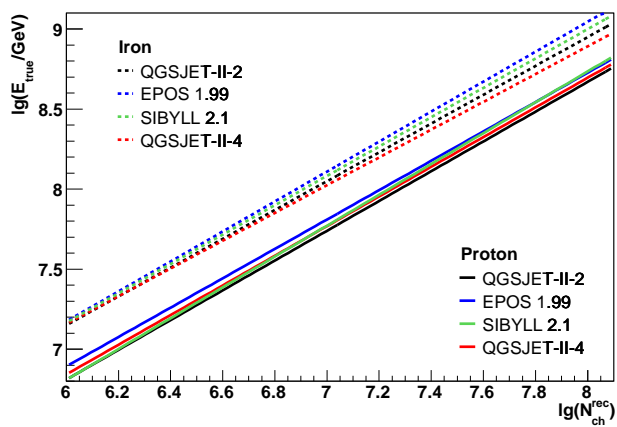}
\caption{Energy calibration functions for assumed pure proton and iron primaries for
  the observable $N_{ch}$ for different interaction models.}
\label{fig3}
\end{figure}

For the estimation of the primary energy, the first step is that
the shower size per individual event is corrected for attenuations in the
atmosphere by the Constant Intensity Cut (CIC) method.
To determine the correlation between the number of charged particles 
and the primary energy, Monte-Carlo simulations were used then,
based on different hadronic interaction models.
The correlation of the primary energy as a function of the number of charged
particles is plotted in Fig.\ \ref{fig3} 
for the assumption of primary protons and iron nuclei, respectively, 
as well as for the different interaction models.
Assuming a linear dependence in logarithmic scale:
lg$E$ = $a + b\cdot$lg($N_{ch}$) and a primary composition,
the linear fit is applied in the range of full efficiencies.
The energy calibration depends on the hadronic interaction models,
so that the fittings are performed individually and
the resulting coefficients of the energy calibration for QGSJET-II-4 are
$a = 1.28 \pm 0.32$ and $b = 0.93 \pm 0.03$,
and $a = 1.95 \pm 0.22$ and $b = 0.87 \pm 0.03$
with a reduced $\chi^{2}$ of 1.27 for proton and 0.88 for iron, respectively.
The fit results of QGSJET-II-2, EPOS 1.99 and SIBYLL 2.1 models 
are summarized in Ref.\ \cite{bib:Kang}.

\begin{figure}
\includegraphics[width=90mm]{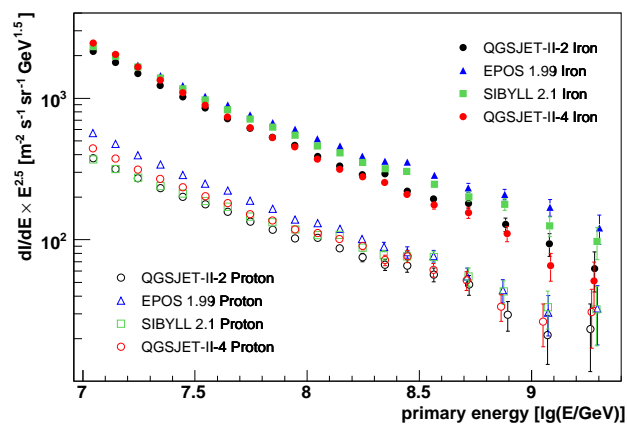}
\caption{
  Reconstructed all-particle energy spectra from KASCADE-Grande shower
  size for assuming proton and iron composition, based on different
  hadronic interaction models of QGSJET-II-2, EPOS 1.99, SIBYLL 2.1 and QGSJET-II-4.}
\label{fig4}
\end{figure}

Figure \ref{fig4} presents the resulting all-particle energy spectra 
obtained after applying the energy reconstruction functions,
based on the assumption of iron and proton for QGSJET-II-4, 
together with the results for QGSJET-II-2, EPOS 1.99 and SIBYLL 2.1 models,
where the shower to shower fluctuations were not properly taken into account yet.

Assuming the iron showers, the spectrum of the QGSJET-II-4 model tends to be close to the one of QGSJET-II-2.
The spectral slopes of all four models show a slight discrepancy over the whole energy range.
It is because of the different ratio of $N_{ch}/N_{\mu}$ of the different hadronic interaction models,
so that the total fluxes are shifted.
However, all the spectra show a similar feature, as well as a similar tendency concerning the assumption of primary masses.   
In addition, the resulting all-particle energy spectra of four different interaction models show 
that they cannot be described by a single power law.
Such thing could imply possibly different elemental composition in the
transition region from galactic to extragalactic origin of cosmic rays.

The total systematic uncertainty of QGSJET-II-2 for proton and iron is
21\% and 10\%, respectively, at the primary energy of 10$^{17}$ eV. 
The estimations of systematic uncertainties for QGSJET-II-4,
are currently being performed. 
It is, however, expected to be about the same order of other models.

\begin{figure}[t]
\begin{center}
\includegraphics[width=90mm]{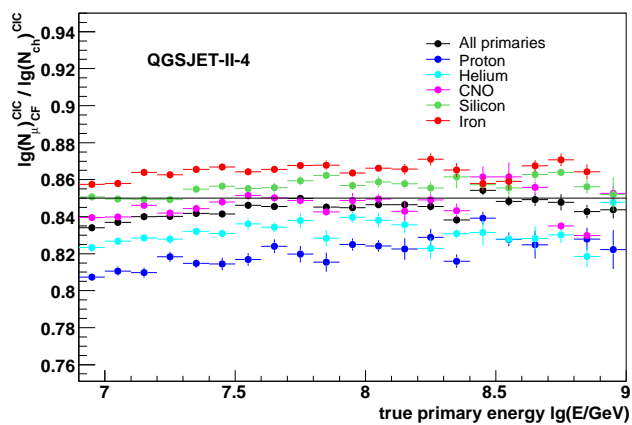}
\caption{The shower size ratio of $Y_{CIC} = lg (N_{\mu}) / lg (N_{ch})$ as a function 
of the true primary energy for the QGSJET-II-4 model.}
\label{fig5}
\end{center}
\end{figure}

\begin{center}
\begin{table*}[]
\hfill{}
\begin{tabular}{lccccc}
\hline
 & $lg(E_{k}$/GeV) & $\gamma_{1}$ & $\gamma_{2}$ & $\Delta \gamma$ & $\chi^{2}$/ndf\\ 
\hline
QGSJET-II-2 & 7.76 $\pm$ 0.06 & 2.92 $\pm$ 0.01 & 3.11 $\pm$ 0.03 & 0.19 & 0.69\\
SIBYLL 2.1  & 7.75 $\pm$ 0.09 & 2.87 $\pm$ 0.03 & 3.15 $\pm$ 0.05 & 0.28 & 1.28\\
EPOS 1.99   & 7.71 $\pm$ 0.06 & 2.76 $\pm$ 0.03 & 3.18 $\pm$ 0.06 & 0.42 & 0.98\\
QGSJET-II-4 & 7.73 $\pm$ 0.14 & 2.88 $\pm$ 0.03 & 3.18 $\pm$ 0.04 & 0.30 & 0.96\\ 
\hline
\end{tabular}
\hfill{}
\caption{The breaking positions and the spectral slopes 
after applying a broken power law fit to the spectra of electron-poor.}
\label{table}
\end{table*}
\end{center}

\section{Spectra of individual mass groups}
Air showers induced by heavier primary particles develop earlier in the atmosphere 
due to their larger cross section for interacting with air nuclei,
and produce relatively larger muon numbers at ground level. 
Therefore, the fraction of muons to the all charged particles at observation level
characterize the mass of the primary particles, i.e. electron-rich showers are generated by light primary nuclei
and electron-poor showers by heavy nuclei, respectively.
Since KASCADE-Grande measures the particle numbers well after the shower maximum, 
the measured showers were separated into electron-poor and electron-rich events
representing heavy and light mass groups.
For this method, the shower size ratio of $Y_{CIC} = lg N_{\mu} / lg N_{ch}$ is used 
to separate the events, where $N_{\mu}$ and $N_{ch}$ are the muon and the charged particle numbers
corrected for attenuation effects in the atmosphere by the CIC method.

\begin{figure}
\begin{center}
\includegraphics[width=90mm]{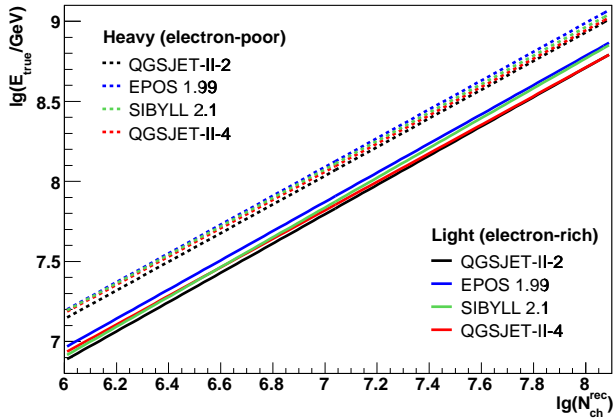}
\caption{
Energy calibration functions for electron-poor and electron-rich
mass groups for different interaction models.}
\label{fig6}
\end{center}
\end{figure}

\begin{figure}
\begin{center}
\includegraphics[width=90mm]{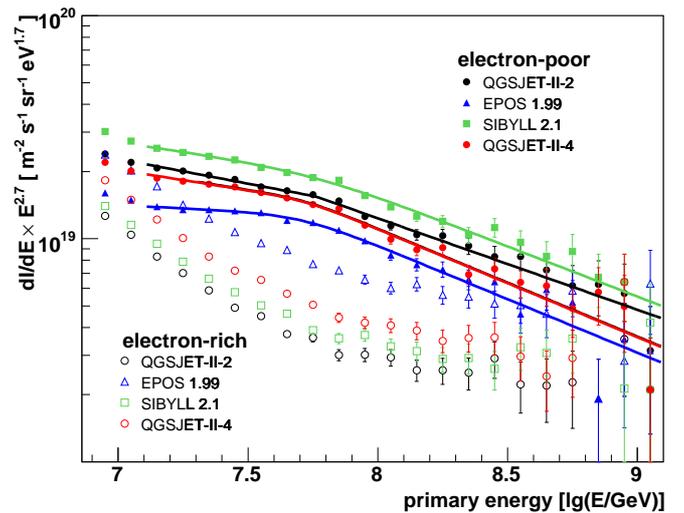}
\caption{Reconstructed energy spectra of the electron-poor and electron-rich
  components, based on different
  hadronic interaction models of QGSJET-II-2, EPOS 1.99, SIBYLL 2.1 and QGSJET-II-4.
The lines show the applied broken power law fits.}
\label{fig7}
\end{center}
\end{figure}

In Fig. \ref{fig5}, the ratio of the shower size as a function of the true primary energy for five different primaries
is plotted.
For the QGSJET-II-4 model, the optimal separation value in between electron-rich and electron-poor
mass groups is $Y_{CIC}$ = 0.85, while it is 0.84 for QGSJET-II-2 and SIBYLL 2.1, and 0.86 for EPOS 1.99.
I.e. the events larger than the value of 0.85 are taken as heavy mass group into account, 
whereas events smaller than 0.85 as light mass group.
After applying the $Y_{CIC}$ selection, 
the energy spectra of electron-rich and electron-poor are reconstructed
only by using the shower size, where
the energy is calibrated by simulations based on the four different hadronic models, shown in Fig. \ref{fig6}.
Figure \ref{fig7} shows the the reconstructed energy spectra of heavy and light mass groups.
Performing the fit of a broken power law, 
the breaking positions and the spectral slopes are summarized in Table \ref{table}.
In the spectra of the heavy primaries, i.e. electron-poor events, a clear knee-like feature
can be seen at just below 10$^{17}$ eV for all four different models.
A remarkable hardening feature above 10$^{17}$ eV in the spectrum of the light primaries 
is observed as well in all hadronic interaction models.

\section{Conclusions}
Based on simulations with the different hadronic interaction models of
QGSJET-II-2, EPOS 1.99, SIBYLL 2.1 and QGSJET-II-4, their influences
on the reconstructed all-particle energy spectrum are investigated 
by means of the shower size measurements of the charged particle component
measured by KASCADE-Grande.
For the all-particle energy spectrum, 
the spectral shapes and structures are in reasonable agreement among four interaction models.
Even if the different hadronic models would give some different values of $Y_{CIC}$,
the spectral shapes of the resulting energy spectra of these mass components 
present a similar tendency for four different interaction models.
Moreover, this result is consistent with another KASCADE-Grande analysis based on
different observables.
In the KASCADE-Grande measurements, we observed similar structures of the all-particle energy spectrum, 
as well as the spectra of the heavy and light mass components,
for different hadronic interaction models.

\vspace*{0.5cm}
\footnotesize{{\bf Acknowledgment:}{
The authors would like to thank the members of the engineering 
and technical staff of the KASCADE-Grande collaboration, who 
contributes to the success of the experiment. 
The KASCADE-Grande experiment was supported in Germany
by the BMBF and by the Helmholtz Alliance for Astroparticle
Physics - HAP funded by the Initiative and Networking
Fund of the Helmholtz Association,
by the MIUR and INAF of Italy,
the Polish Ministry of Science and Higher Education, and
the Romanian Authority for Scientific Research UEFISCDI 
(PNII-IDEI grants 271/2011 and 17/2011).}}

\end{document}